\documentclass[a4paper]{jpconf}
\usepackage{graphicx}
\usepackage{amsbsy}

\def\mprl{\mbox{\tiny $\|$}}

\newcommand{\D}{\mathrm{d}} % differential
\begin{document}
\title{Neutrino processes $\nu \bar\nu \to e^- e^+$ and $\nu \to \nu e^- e^+$ in~a~strong magnetic field}

\author{A V Kuznetsov$^1$, D A Rumyantsev$^{1,\ast}$ and V N Savin$^{2,\ast \ast}$}
\address{$^1$ Division of Theoretical Physics, Department of Physics, 
Yaroslavl State P G Demidov University, Sovietskaya 14,
150000 Yaroslavl, Russia}
\address{$^2$ Division of Physics, Yaroslavl Higher Military School of Air Defence, Moskovskiy Prosp. 28, 
150001 Yaroslavl, Russia}

\ead{avkuzn@uniyar.ac.ru, $^{\ast}$rda@uniyar.ac.ru, $^{\ast \ast}$vs\_post07@mail.ru}

\begin{abstract}
The processes of neutrino production of electron--positron pairs, $\nu \bar\nu \to e^- e^+$ and $\nu \to \nu e^- e^+$, in a magnetic field of arbitrary strength, where electrons and positrons can be created in the states corresponding to excited Landau levels, are analysed. The results can be applied for calculating the efficiency of the electron--positron plasma production by neutrinos in the conditions of the Kerr black hole accretion disc considered by experts as the most possible source of a short cosmological gamma burst. 
\end{abstract}

%%%%%%%%%%%%%%%%%%%%%%%%%%%%%%%%%%%%%%%%%%%%%%%%%%
\section{Introduction}
%%%%%%%%%%%%%%%%%%%%%%%%%%%%%%%%%%%%%%%%%%%%%%%%%%

An intense electromagnetic field makes possible the processes which are forbidden in a vacuum such as 
the neutrino production of an electron--positron pair $\nu \to \nu e^- e^+$. 
The list of papers devoted to an analysis of this process and the collection of the results obtained 
could be found e.g. in~\cite{KM_Book_2013}. 
In most cases, calculations of this kind were made either in the crossed field approximation, 
or in the limit of a superstrong field much greater than the critical value of 
$B_e = m_e^2/e \simeq 4.41\times 10^{13}$~G (we use natural units
$c = \hbar = k_{\rm{B}} = 1$), 
when the electrons and positrons are born
in states corresponding to the ground Landau level. 
However, there exist physical situations of the so-called moderately strong magnetic 
field, 
$p_\perp^2 \ge e B \gg m_e^2$, when electrons and positrons mainly occupy the ground
Landau level, however, a noticeable fraction may be produced at the next levels. 

The indicated hierarchy of physical parameters corresponds to the conditions of the Kerr black hole accretion disk, regarded by experts as the most likely source of a short cosmological gamma-ray burst. 
The disc is a source of copious neutrinos and anti-neutrinos, which partially annihilate above the disc
and turn into $e^{\mp}$ pairs, $\nu \bar\nu \to e^- e^+$. This process was proposed and investigated in many details 
(for the list of references see e.g.~\cite{Beloborodov:2011,Kuznetsov:2014}) 
as a possible mechanism for creating relativistic, $e^{\mp}$-dominated jets that could power observed 
gamma-ray bursts. 
In~\cite{Beloborodov:2011}, in addition to $\nu \bar\nu$ annihilation, the contribution
of the magnetic field-induced process $\nu \to \nu e^- e^+$ to the neutrino energy deposition rate around
the black hole was also included for the first time.
The authors~\cite{Beloborodov:2011} concluded in part, that the process $\nu \to \nu e^- e^+$ could dominate over 
the basic process $\nu \bar\nu \to e^- e^+$.
They used the result for the energy deposition rate in the process $\nu \to \nu e^- e^+$ 
obtained in~\cite{Kuznetsov:1997a,Kuznetsov:1997b} in the crossed field limit, while in those physical conditions ($B$ to 180~$B_e$, $E_\nu$ to 25 MeV) the approximation of a crossed field is poorly applicable 
(as well as the approximation of a superstrong field when $e^-e^+$ are created in the ground Landau level).
The next Landau levels can be also excited, as we have shown in our paper~\cite{Kuznetsov:2014}.
Furthermore, the authors~\cite{Beloborodov:2011} considered the process $\nu \bar\nu \to e^- e^+$ 
without taking account of the magnetic field influence. 

Thus, the aim of this paper is the study of the processes $\nu \bar\nu \to e^- e^+$ and $\nu \to \nu e^- e^+$ 
in the physical conditions of the moderately strong 
magnetic field, where the electrons and positrons would be born in the states corresponding to the excited
Landau levels. 
Possible astrophysical applications are discussed.

%%%%%%%%%%%%%%%%%%%%%%%%%%%%%%%%%%%%%%%%%%%%%%%%%%
\section{Neutrino process $\nu \to \nu e^- e^+$ in~a~strong magnetic field}
%%%%%%%%%%%%%%%%%%%%%%%%%%%%%%%%%%%%%%%%%%%%%%%%%%

The total probability of the process $\nu \to \nu e^-_{(n)} e^+_{(\ell)}$, when the electron and the positron are created 
in the $n$th and $\ell$th Landau levels, 
is, in a general case, the sum of the probabilities of the four polarization channels:
\begin{equation}
\label{eq:Wtot}
W_{n \ell} = W^{--}_{n \ell} + W^{-+}_{n \ell} + W^{+-}_{n \ell} + W^{++}_{n \ell} \, .
\end{equation}
For each of the channels, the differential probability over the final neutrino momentum 
per unit time, after integration over the momenta of the electron and positron, is reduced to one
nontrivial integral:
\begin{eqnarray}
\D W^{s s'}_{n \ell} = 
\frac{\beta \, \D^3 P'}{(2 \pi)^4 16 E E'} \, 
\int \, 
\frac{\D p_z}{\varepsilon_n \, \varepsilon'_{\ell}} \, \delta(\varepsilon_n + \varepsilon'_{\ell} - q_0) \,
|{\cal M}_{n \ell}^{s s'}|^2 \, ,
\label{eq:dw2} 
\end{eqnarray}
where $\varepsilon_n = \sqrt{M_n^2 + p_z^2}$, $M_n = \sqrt{m_e^2 + 2 \beta n}$, $\beta = e B$. 
The energy of the initial
neutrino should exceed a certain threshold value.
In the reference frame, 
where the momentum of the initial neutrino directed at an angle
$\theta$ to the magnetic field, the threshold energy is given by:
\begin{equation}
E \, \sin \theta \ge M_{n} + M_{\ell} \, .
\label{eq:condE} 
\end{equation}
Some details of calculations can be found in our paper~\cite{Kuznetsov:2014}.

The probability of the $\nu \to \nu e^- e^+$ process defines 
its partial contribution into the neutrino opacity of the medium. 
The estimation of the neutrino mean free 
path with respect to this process gives the result which is too large~\cite{KM_Book_2013} 
compared with the typical size of any compact astrophysical object, 
where a strong magnetic field could exist. 
However, a mean free path does not exhaust the neutrino physics in 
a medium. In astrophysical applications, we could consider 
the values that probably are more essential, namely, the mean values 
of the neutrino energy and momentum losses, caused by the influence of an external magnetic field. 
These values can be described by the four-vector of losses $Q^{\alpha}$, 
\begin{equation}
Q^\alpha \, = \, E \int q^\alpha \, \D W = - E \, ({\cal I}, {\bf F}) \,.
\label{eq:Q0}
\end{equation}
where $q$ is the difference of the momenta of the initial and final neutrinos, 
$q = P - P'$, $\D W$ is the total differential probability of the process. 
The zeroth component of $Q^{\alpha}$ is connected with the mean energy lost 
by a neutrino per unit time due to the process considered, 
${\cal I} = \D E/\D t$.
The space components of the four-vector~(\ref{eq:Q0}) are similarly 
connected with the mean neutrino momentum loss per unit time, 
${\bf F} = \D {\bf P}/\D t$. 
It should be noted that the four-vector of losses $Q^{\alpha}$ can be used for evaluating 
the integral effect of neutrinos on plasma in the conditions of not very dense plasma, 
where an one-interaction approximation of a neutrino with plasma is valid.

In~\cite{Beloborodov:2011}, the formula for the energy deposition rate was taken, which 
was calculated in the crossed field limit~\cite{Kuznetsov:1997a,Kuznetsov:1997b}.  
However, in the region of the physical parameters 
used in~\cite{Beloborodov:2011} ($B$ to 180 $B_e$, $E_\nu$ to 25 MeV), the approximation of a crossed field is poorly applicable,
as well as the approximation of a superstrong field when $e^- e^+$ are created in the ground Landau level.
The contribution of the next Landau levels which can be also excited, should be taken into account.
In~\cite{Kuznetsov:2014}, the results 
are presented of our calculation of the mean neutrino energy losses caused by
the process $\nu \to \nu e^- e^+$ in a moderately strong magnetic field, i.e. in the conditions 
of the Kerr black hole accretion disk.  
It was shown that the crossed field limit gives the overstated result which is in orders of magnitude greater 
than the sum of the contributions of lower excited Landau levels.   
On the other hand, the results with $e^- e^+$ created at the ground Landau level give the main contribution to 
the energy deposition rate, and almost exhaust it at $B = 180 B_e$. 

This would mean that the conclusion~\cite{Beloborodov:2011} that the contribution of the process 
$\nu \to \nu e^- e^+$ to the efficiency of the electron-positron plasma production by neutrino exceeds
the contribution of the annihilation channel $\nu \bar\nu \to e^- e^+$, and that the first process 
dominates the energy deposition rate, does not have a sufficient basis.  
A new analysis of the efficiency of energy deposition by neutrinos through both processes, $\nu \bar\nu \to e^- e^+$ 
and $\nu \to \nu e^- e^+$, in a hyper-accretion disc around a black hole should be performed, with taking 
account of our results~\cite{Kuznetsov:2014} for the process $\nu \to \nu e^- e^+$. 

%%%%%%%%%%%%%%%%%%%%%%%%%%%%%%%%%%%%%%%%%%%%%%%%%%
\section{The strong magnetic field influence on the process $\nu \bar\nu \to e^- e^+$}
%%%%%%%%%%%%%%%%%%%%%%%%%%%%%%%%%%%%%%%%%%%%%%%%%%

The local energy-momentum 
deposition rate due to the process $\nu \bar\nu \to e^- e^+$ is defined by the equation~\cite{Birkl:2007}:
\begin{equation}
Q^\alpha_{\nu \bar\nu} =
\int \frac{\D^3 p}{(2 \pi)^3} \, f_{\nu} (p) \int \frac{\D^3 p'}{(2 \pi)^3} \, f_{\bar\nu} (p') \,
 \left( p^\alpha + p'^\alpha \right) \frac{(p p')}{E E'} \, 
\sigma (\nu \bar\nu \to e^- e^+) \,,
\label{eq:Q1}
\end{equation}
where $\sigma$ is the cross-section of the process, 
$p$ and $p'$ are the four-momenta of the neutrino and antineutrino,
$f_{\nu} (p)$ and $f_{\bar\nu} (p')$ are the local distribution functions depending 
on the distribution functions at the surface of
the black hole accretion disc, and on the details of propagation.

In a strong magnetic field, the cross-section takes the form:
\begin{equation}
\sigma (\nu \bar\nu \to e^- e^+)  = \sum\limits_{f=e,\mu,\tau} \, \sum\limits_{n, \ell} \, 
\sigma (\nu_f {\bar\nu}_f \to  e^-_{(n)} e^+_{(\ell)}) \,,
\label{eq:sigmaB}
\end{equation}
where the upper limit of summation over $n, \ell$ is defined by the condition 
$(M_n + M_{\ell})^2 \le ( p + p' )_{\mprl}^2$, ($q_{\mprl}^2=q_0^2-q_z^2$, if $z$ is along $\bf B$). 

\begin{figure}[!t] 
\begin{center}
\includegraphics*[width=0.9\textwidth]{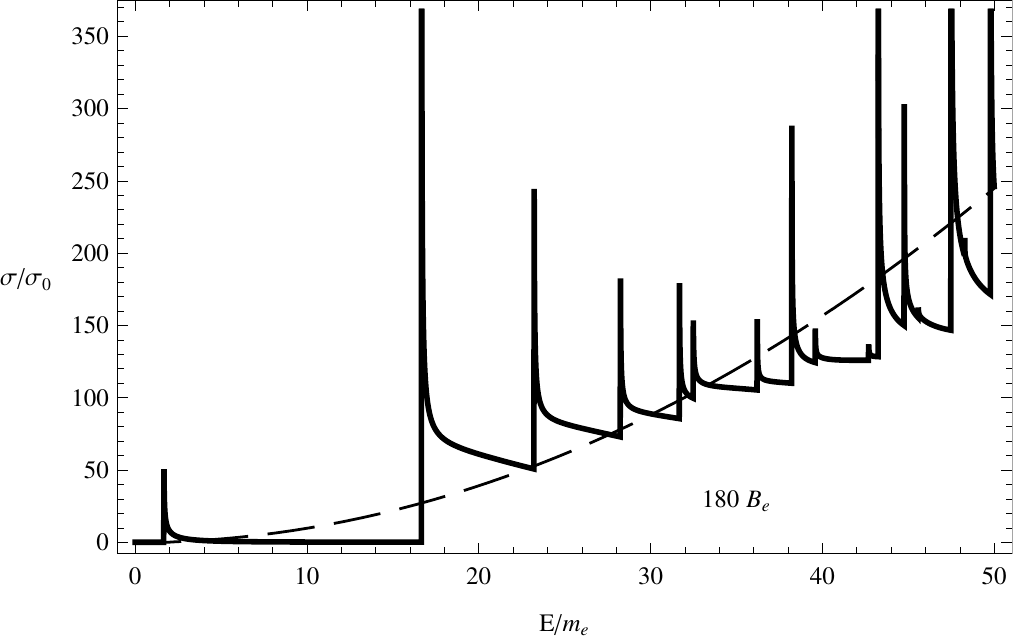}
\end{center}
\caption{Cross-section of the process $\nu \bar\nu \to e^- e^+$ in the case $E=E'$ and for fixed angles: 
the energy dependence in the field (solid line) and in vacuum (dashed line); 
$\sigma_0 = {4 \, G_{\mathrm{F}}^2 \, m_e^2}/{\pi}$ is the so-called typical weak cross-section.}
\label{fig:function180}
\end{figure}

Unlike the cross-section in vacuum where it depends on the Mandelstam parameter $S$ only, 
the cross-section in a magnetic field depends on the set of kinematic variables, e.g.: energies $E, E'$, two polar angles and one azimuth angle. In figure~\ref{fig:function180}, we take for the sake of illustration the case $E=E'$, and take certain angles. The dependence is presented of $\sigma$ (solid line) and $\sigma_{\rm vac}$ (dashed line) on $E$.
The cross-section has a peculiar ``sawtooth'' profile due to the square-root singularities~\cite{Klepikov:1954}, 
which is similar 
to the profile of the process $\gamma^\ast \to e^- e^+$ width in a strong field~\cite{Daugherty:1983, Baier:2007}. 
After averaging over small intervals $E \pm \Delta E$, the dependence becomes smoother. 
It can be seen that in calculations of the energy-momentum deposition rate by integration 
over the neutrino and antineutrino momenta, the field influence appears to be inessential.

%---------------------------------------------------------------------------------------------
%---------------------------------------------------------------------------------------------

%%%%%%%%%%%%%%%%%%%%%%%%%%%%%%%%%%%%%%%%%%%%%%%%%%
\section{Conclusions}
%%%%%%%%%%%%%%%%%%%%%%%%%%%%%%%%%%%%%%%%%%%%%%%%%%

\begin{itemize}
\item
The processes $\nu \to \nu e^- e^+$ and $\nu \bar\nu \to e^- e^+$ are investigated in the magnetic field of an arbitrary strength, when $e^- e^+$
can be produced in the excited Landau levels.  
\item
The neutrino energy losses due to the process $\nu \to \nu e^- e^+$ are calculated. The results should be used for calculations of the efficiency of the $e^- e^+$ plasma production by neutrinos in 
the conditions of the Kerr black hole accretion disk. In these conditions, the crossed field limit gives the overstated result which is in orders of magnitude greater 
than the sum over the lower Landau levels. 
\item
The cross-section of the process $\nu \bar\nu \to e^- e^+$ in a strong field, has a peculiar ``sawtooth'' profile, which is close to the vacuum cross-section after averaging. In calculations of the energy-momentum deposition rate by integration over the neutrino and antineutrino momenta, the field influence appears to be inessential. 
\end{itemize}

\ack
The study was performed with the support by the Project No.~92 within the base part of the State Assignment 
for the Yaroslavl University Scientific Research, and was supported in part by the 
Russian Foundation for Basic Research (Project No. \mbox{14-02-00233-a}).


\section*{References}
\begin{thebibliography}{9}
\bibitem{KM_Book_2013}
   Kuznetsov A and Mikheev N 2013 {\it Electroweak Processes in External Active Media} 
   (Berlin, Heidelberg: Springer-Verlag)

\bibitem{Beloborodov:2011} 
Zalamea I and Beloborodov A M 2011 {\it Mon. Not. R. Astron. Soc.} {\bf 410} 2302

\bibitem{Kuznetsov:2014}
Kuznetsov A V, Rumyantsev D A and Savin V N 2014 
{\it Int. J. Mod. Phys.} A {\bf 29} 1450136% ({\it Preprint} arXiv:1406.3904[hep-ph])

\bibitem{Kuznetsov:1997a}
   Kuznetsov A V and Mikheev N V 1997 {\it Phys. Lett.} B {\bf 394} 123
 
\bibitem{Kuznetsov:1997b}
   Kuznetsov A V and Mikheev N V 1997 
   {\it Phys. At. Nucl.} {\bf 60} 1865 (Original Russian text: {\it Yad. Fiz.} {\bf 60} 2038)

\bibitem{Birkl:2007}
   Birkl R, Aloy M A, Janka H-Th and M{\" u}ller E 2007
  {\it Astron. Astrophys.} {\bf 463} 51

\bibitem{Klepikov:1954}
  Klepikov N P 1954 (In Russian) {\it Zh. Eksp. Teor. Fiz.} {\bf 26} 19 

\bibitem{Daugherty:1983}
  Daugherty J K and Harding A K 1983 
  {\it Astrophys. J.} {\bf 273} 761

\bibitem{Baier:2007}
Baier V N and Katkov V M 2007 {\it Phys. Rev.} D {\bf 75} 073009

\end{thebibliography}
\end{document}